\documentclass[11pt]{article}

\usepackage[bookmarks=false,colorlinks]{hyperref}

\pdfoutput=1

\usepackage{amsmath}
\usepackage{amssymb}
\usepackage{algorithm}
\usepackage{algorithmic}
\usepackage{bm}
\usepackage{bbm}
\usepackage[numbers]{natbib}
\usepackage{graphicx}
\usepackage[margin=0.8in]{geometry}
\usepackage{authblk}

\floatstyle{plain}
\newfloat{myalgo}{tbhp}{mya}

\newenvironment{Algorithm}[2][tbh]%
{\begin{myalgo}[#1]
\centering
\begin{minipage}{#2}
\begin{algorithm}[H]}%
{\end{algorithm}
\end{minipage}
\end{myalgo}}

\newcommand{\pd}[2]{ \frac{\partial #1}{\partial #2} }
\newcommand{\comment}[1]{}

\title{\Large \bfseries Information-geometric Markov Chain Monte Carlo methods using Diffusions}

\author[1]{Samuel Livingstone \footnote{Corresponding author: samuel.livingstone@ucl.ac.uk} }
\author[2]{Mark Girolami}
\affil[1]{ \normalsize Department of Statistical Science, University College London, Gower Street, London, UK}
\affil[2]{ \normalsize Department of Statistics, University of Warwick, Coventry, UK}

\begin{document}
\maketitle

\abstract{Recent work incorporating geometric ideas in Markov chain Monte Carlo is reviewed in order to highlight these advances and their possible application in a range of domains beyond Statistics.  A full exposition of Markov chains and their use in Monte Carlo simulation for Statistical inference and molecular dynamics is provided, with particular emphasis on methods based on Langevin diffusions.  After this geometric concepts in Markov chain Monte Carlo are introduced.  A full derivation of the Langevin diffusion on a Riemannian manifold is given, together with a discussion of appropriate Riemannian metric choice for different problems.  A survey of applications is provided, and some open questions are discussed.}


\section{Introduction}

There are three objectives to this article.  The first is to introduce geometric concepts that have recently been employed in Monte Carlo methods based on Markov chains \cite{girolami2011riemann} to a wider audience.  The second is to clarify what a `diffusion on a manifold' is, and how this relates to a diffusion defined on Euclidean space.  Finally we review the state of the art in the field, and suggest avenues for further research.

The connections between some Monte Carlo methods commonly used in Statistics, Physics and application domains such as Econometrics, and ideas from both Riemannian and Information geometry \cite{amari2007methods,marriott2000applications} were highlighted by Girolami \& Calderhead \cite{girolami2011riemann}, and the potential benefits demonstrated empirically.  Two Markov chain Monte Carlo methods were introduced, the manifold Metropolis-adjusted Langevin algorithm and Riemannian manifold Hamiltonian Monte Carlo.  Here we focus on the former for two reasons.  First, the intuition for why geometric ideas can improve standard algorithms is the same in both cases.  Second, the foundations of the methods are quite different, and since the focus of the article is on using geometric ideas to improve performance, we considered a detailed description of both to be unnecessary.  It should be noted, however, that impressive empirical evidence exists for using Hamiltonian methods in some scenarios (e.g. \cite{betancourt2013hamiltonian}).  We refer interested readers to \cite{neal2011mcmc,betancourt2011geometry}.

We take an expository approach, providing a review of some necessary preliminaries from Markov chain Monte Carlo, diffusion processes and Riemannian geometry.  We assume only a minimal familiarity with measure-theoretic probability.  More informed readers may prefer to skip these sections.  We then provide a full derivation of the Langevin diffusion on a Riemannian manifold, and offer some intuition for how to think about such a process.  We conclude Section \ref{sec:geometry} by presenting the Metropolis-adjusted Langevin algorithm on a Riemannian manifold.

A key challenge in the geometric approach is which manifold to choose.  We discuss this in Subsection \ref{sub:metric}, and review some candidates that have been suggested in the literature, along with the reasoning for each.  Rather than provide a simulation study here, we instead reference studies where the methods we describe have been applied in Section \ref{sec:applications}.  In Section \ref{sec:discussion} we discuss several open questions which we feel could be interesting areas of further research, and of interest to both theorists and practitioners.

Throughout $\pi(\cdot)$ will refer to an $n$-dimensional probability distribution, and $\pi(x)$ its density with respect to Lebesgue measure.



\section{Markov Chain Monte Carlo} \label{sec:mcmc}

Markov chain Monte Carlo (MCMC) is a set of methods for drawing samples from a distribution $\pi(\cdot)$ defined on a measurable space $(\mathcal{X}, \mathcal{B})$ whose density is only known up to some proportionality constant.  Although the $i$th sample is dependent on the $(i-1)$th, the Ergodic Theorem ensures that for an appropriately constructed Markov chain  with invariant distribution $\pi(\cdot)$, long-run averages are consistent estimators for expectations under $\pi(\cdot)$.  As a result, MCMC methods have proven useful in Bayesian Statistical inference, where often the posterior density $\pi(x|y) \propto f(y|x)\pi_0(x)$ for some parameter $x$ is only known up to a constant \cite{robert2004monte}.  Here we briefly introduce some concepts from general state space Markov chain theory together with a short overview of MCMC methods.  The exposition follows \cite{tierney1994markov}.

\subsection{Markov Chain Preliminaries}

A time-homogeneous Markov chain $\{ X_m \}_{m \in \mathbb{N}}$ is a collection of random variables $X_m$ each of which is defined on a measurable space $(\mathcal{X},\mathcal{B})$, such that:
\begin{equation}
\mathbb{P}[X_m \in A | X_0 = x_0,...,X_{m-1}=x_{m-1}] = \mathbb{P}[X_m \in A | X_{m-1}=x_{m-1}],
\end{equation}
for any $A \in \mathcal{B}$.  We define the transition kernel $P(x_{m-1},A) = \mathbb{P}[X_m \in A | X_{m-1}=x_{m-1}]$ for the chain to be a map for which $P(x,\cdot)$ defines a distribution over $(\mathcal{X},\mathcal{B})$ for any $x \in \mathcal{X}$ and $P(\cdot,A)$ is measurable for any $A \in \mathcal{B}$.  Intuitively, $P$ defines a map from points to distributions in $\mathcal{X}$.  Similarly we define the $m$-step transition kernel to be
\begin{equation}
P^m(x_0,A) = \mathbb{P}[X_m \in A | X_0 = x_0].
\end{equation}
We call a distribution $\pi(\cdot)$ \emph{invariant} for $\{ X_m \}_{m \in \mathbb{N}}$ if
\begin{equation} \label{eqn:stat1}
\pi(A) = \int_{\mathcal{X}} P(x,A) \pi(dx) 
\end{equation}
for all $A \in \mathcal{B}$.  If $P(x,\cdot)$ admits a density $p(x'|x)$, this can be equivalently written:
\begin{equation} \label{eqn:stat2}
\pi(x') = \int_{\mathcal{X}} \pi(x) p(x'|x)dx.
\end{equation}
The connotation of (\ref{eqn:stat1}) and (\ref{eqn:stat2}) is that if $X_m \sim \pi(\cdot)$, then $X_{m+s} \sim \pi(\cdot)$ for any $s \in \mathbb{N}$.  In this instance we say the chain is `at stationarity'.  Of interest to us will be Markov chains for which there is a unique invariant distribution which is also the \emph{limiting} distribution for the chain, meaning that for any $x_0 \in \mathcal{X}$ for which $\pi(x_0) > 0$
\begin{equation} \label{eqn:equil}
\lim_{m \to \infty} P^m(x_0,A) = \pi(A)
\end{equation}
for any $A \in \mathcal{B}$.  Certain conditions are required for (\ref{eqn:equil}) to hold, but for all Markov chains presented here these are satisfied (though see \cite{tierney1994markov}).

A useful condition which is sufficient (though not necessary) for $\pi(\cdot)$ to be an invariant distribution is \emph{reversibility}, which can be shown by the relation
\begin{equation} \label{eqn:reverse}
\pi(x)p(x'|x) = \pi(x')p(x|x').
\end{equation}
Integrating over both sides with respect to $x$ we recover (\ref{eqn:stat2}).  In words, a chain is reversible if at stationarity the probability that $x_i \in A$ and $x_{i+1} \in B$ is equal to the probability that $x_{i+1} \in A$ and $x_i \in B$.  The relation (\ref{eqn:reverse}) will be the primary tool used to construct Markov chains with a desired invariant distribution in the next section.

\subsubsection{Monte Carlo estimates from Markov chains}

Of most interest here are estimators constructed from a Markov chain.  The Ergodic Theorem states that for any chain $\{ X_m \}_{m \in \mathbb{N}}$ satisfying (\ref{eqn:equil}) and any $g \in L^1(\pi)$ we have that
\begin{equation}
\lim_{m \to \infty} \frac{1}{m} \sum_{i=1}^m g(X_i) = \mathbb{E}_{\pi}[g(X)] 
\end{equation}
with probability one \cite{robert2004monte}.  This is a Markov chain analogue to the Law of Large numbers.

The efficiency of estimators of the form $\hat{t}_m = \sum_i g(X_i) / m$ can be assessed through the \emph{autocorrelation} between elements in the chain.  We will assess the efficiency of $\hat{t}_m$ relative to estimators $\bar{t }_m =  \sum_i g(Z_i) / m$, where $\{ Z_i \}_{m \in \mathbb{N}}$ is a sequence of independent random variables each having distribution $\pi(\cdot)$.  Provided $\text{Var}_{\pi}[g(Z_i)] < \infty$ then $\text{Var}[\bar{t}_m] = \text{Var}_{\pi}[g(Z_i)]/m$.  We now seek a similar result for estimators of the form $\hat{t}_m$.

It follows directly from the Kipnis-Varadhan Theorem \cite{kipnis1986central} that an estimator $\hat{t}_m$ from a reversible Markov chain for which $X_0 \sim \pi(\cdot)$ satisfies:
\begin{equation} \label{eqn:kv2}
\lim_{m \to \infty} \frac{\text{Var}[ \hat{t}_m ]}{\text{Var}[ \bar{t}_m ] } = 1 + 2\sum_{i=1}^{\infty} \rho^{(0,i)} = \tau,
\end{equation} 
provided that $\sum_{i=1}^\infty i|\rho^{(0,i)}|<\infty$, where $\rho^{(0,i)} = \text{Corr}_{\pi}[g(X_0),g(X_i)]$.  We will refer to the constant $\tau$ as the \emph{autocorrelation time} for the chain .

Equation (\ref{eqn:kv2}) implies that for large enough $m$, $\text{Var}[ \hat{t}_m ] \approx \tau \text{Var} [ \bar{t}_m ]$.  In practical applications, the sum in (\ref{eqn:kv2}) is truncated to the first $p-1$ realisations of the chain, where $p$ is the first instance at which $|\rho^{(0,p)}| < \epsilon$ for some $\epsilon > 0$ \cite{plummer2006coda}.

Another commonly used measure of efficiency is the \emph{effective} sample size $m_{eff} = m/\tau$, which gives the number of independent samples from $\pi(\cdot)$ needed to give an equally efficient estimate for $\mathbb{E}_{\pi} [g(X)]$.  Clearly minimising $\tau$ is equivalent to maximising $m_{eff}$.

The measures arising from (\ref{eqn:kv2}) give some intuition for what sort of Markov chain gives rise to efficient estimators.  However, in practice the chain will never be at stationarity.  So we also assess Markov chains according to how far away they are from this point.  For this, we need to measure how close $P^m(x_0,\cdot)$ is from $\pi(\cdot)$, which requires a notion of distance between probability distributions.

Although there are several appropriate choices \cite{gibbs2002choosing}, a common option in the Markov chain literature is the Total variation distance
\begin{equation} \label{eqn:TV}
\| \mu(\cdot) - \nu(\cdot) \|_{TV} := \sup_{A \in \mathcal{B}} |\mu(A) - \nu(A)|,
\end{equation}
which informally gives the largest possible difference between the probabilities of a single event in $\mathcal{B}$ according to $\mu(\cdot)$ and $\nu(\cdot)$.  If both distributions admit densities, (\ref{eqn:TV}) can be written
\begin{equation}
\| \mu(\cdot) - \nu(\cdot) \|_{TV} = \frac{1}{2} \int_{\mathcal{X}} |\mu(x) - \nu(x)| dx.
\end{equation}
which is proportional to the $L_1$ distance between $\mu(x)$ and $\nu(x)$.  Our metric $\|\cdot\|_{TV} \in [0,1]$, with $\|\cdot\|_{TV} = 1$ for distributions with disjoint supports and $\|\mu(\cdot) - \nu(\cdot)\|_{TV} = 0$ implying $\mu(\cdot) \equiv \nu(\cdot)$.

Typically for an unbounded $\mathcal{X}$ the distance $\|P^m(x_0,\cdot) - \pi(\cdot)\|_{TV}$ will depend on $x_0$ for any finite $m$.  So bounds on the distance are often sought via some inequality of the form
\begin{equation} \label{eqn:TV2}
\|P^m(x_0,\cdot) - \pi(\cdot)\|_{TV} \leq V(x_0)f(m),
\end{equation}
where $V: \mathcal{X} \to [1,\infty)$ depends on $x_0$ and is called a \emph{drift} function, and $f: \mathbb{N} \to [0,\infty)$ depends on the number of iterations $m$ (and is often defined such that $f(0) = 1$).

A Markov chain is called \emph{geometrically ergodic} if $f(m) = r^m$ in (\ref{eqn:TV2}) for some $0 < r < 1$.  If in addition to this $V$ is bounded above, the chain is called \emph{uniformly ergodic}.  Intuitively, if either condition holds then the distribution of $X_m$ will converge to $\pi(\cdot)$ geometrically quickly as $m$ grows, and in the uniform case this rate is independent of $x_0$.  As well as providing some (often qualitative if $r$ is unknown) bounds on the convergence rate of a Markov chain, geometric ergodicity implies that a central limit theorem exists for estimators of the form $\hat{t}_m$, so that the shape of the distribution is asymptotically Gaussian.  For more detail on this see \cite{jones2001honest, jones2004markov}.

In practice several approximate methods also exist to assess whether a chain is close enough to stationarity for long-run averages to provide suitable estimators (e.g. \cite{gelman1992inference}).  The MCMC practitioner also uses a variety of visual aids to judge whether an estimate from the chain will be appropriate for his or her needs.

\subsection{Markov Chain Monte Carlo}

Now that we have introduced Markov chains we turn to simulating them.  The objective here is to devise a method for generating a Markov chain which has a desired limiting distribution $\pi(\cdot)$.  In addition we would strive for the convergence rate to be as fast as possible, and the effective sample size to be suitably large relative to the number of iterations.  Of course, the computational cost of performing an iteration is also an important practical consideration.  Ideally any method would also require limited problem-specific alterations, so that practitioners are able to use it with as little knowledge of the inner workings as is practical.

Although other methods exist for constructing chains with a desired limiting distribution, a popular choice is the Metropolis--Hastings algorithm \cite{robert2004monte}.  At iteration $i$, a sample is drawn from some \emph{candidate} transition kernel $Q(x_{i-1},\cdot)$, and then either accepted or rejected (in which case the state of the chain remains $x_{i-1}$).  We focus here on the case where $Q(x_{i-1},\cdot)$ admits a density $q(x'|x_{i-1})$ for all $x_{i-1} \in \mathcal{X}$ (though see \cite{tierney1994markov}).  In this case a single step is shown below (the wedge notation $a \wedge b$ denotes the minimum of $a$ and $b$).
\begin{Algorithm}[ht]{15.5cm}
\caption{Metropolis-Hastings, single iteration}
\begin{algorithmic}
\REQUIRE $x_{i-1}$
\STATE Draw $X' \sim Q(x_{i-1},\cdot)$
\STATE Draw $Z \sim U[0,1]$
\STATE Set $\alpha(x_{i-1},x') \leftarrow 1 \wedge \frac{\pi(x')q(x_{i-1}|x')}{\pi(x_{i-1})q(x'|x_{i-1})}$
\IF{$z < \alpha(x_{i-1},x')$}
\STATE Set $x_i \leftarrow x'$
\ELSE
\STATE Set $x_i \leftarrow x_{i-1}$
\ENDIF
\end{algorithmic}
\end{Algorithm}

\vspace{0.5cm}

The `acceptance rate' $\alpha(x_{i-1},x')$ governs the behaviour of the chain.  If it is typically close to one then many proposed moves are accepted and so the current value in the chain is constantly changing.  If it is on average close to zero then many proposals are rejected so the chain will remain in the same place for many iterations.  However, $\alpha \approx 1$ is typically not ideal, often resulting in a large autocorrelation time (see below).  The challenge in practice is to find the right acceptance rate to balance these two extremes.  

Combining the `proposal' and `acceptance' steps, the transition kernel for the resulting Markov chain is
\begin{equation}
P(x,A) = r(x)\delta_x(A) + \int_A \alpha(x,x')q(x'|x)dx',
\end{equation}
for any $A \in \mathcal{B}$, where
\[
r(x) = 1- \int_\mathcal{X} \alpha(x,x')q(x'|x)dx'
\]
is the average probability that a draw from $Q(x,\cdot)$ will be rejected, and $\delta_x(A) = 1$ if $x \in A$ and zero otherwise.  A Markov chain defined in this way will have $\pi(\cdot)$ as an invariant distribution, since the chain is reversible for $\pi(\cdot)$.  We note here that
\begin{align*}
\pi(x_{i-1})q(x_i|x_{i-1})\alpha(x_{i-1},x_i) &= \pi(x_{i-1})q(x_i|x_{i-1}) \wedge \pi(x_i)q(x_{i-1}|x_i) \\
&= \alpha(x_i,x_{i-1})q(x_{i-1}|x_i)\pi(x_i)
\end{align*}
in the case that the proposed move is accepted, and that if the proposed move is rejected then $x_i = x_{i-1}$ so the chain is reversible for $\pi(\cdot)$.  It can be shown that $\pi(\cdot)$ is also the limiting distribution for the chain \cite{robert2004monte}.

The convergence rate and autocorrelation time of a chain produced by the algorithm are dependent on both the choice of proposal $Q(x_{i-1},\cdot)$ and the target distribution $\pi(\cdot)$.  For simple forms of the latter, less consideration is required when choosing the former.  A broad objective among researchers in the field is to find classes of proposal kernels that produce chains which converge and mix quickly for a large class of target distributions.  We first review a simple choice before discussing one which is more sophisticated, and will be the focus of the rest of the article.

\subsection{Random walk proposals}

An extremely simple choice for $Q(x,\cdot)$ is one for which:
\begin{equation} \label{eqn:RWM}
q(x'|x) = q(\|x' - x\|)
\end{equation}
where $\|\cdot\|$ denotes some appropriate norm on $\mathcal{X}$, meaning the proposal is \emph{symmetric}.  In this case, the acceptance rate reduces to:
\begin{equation} \label{eqn:RWM2}
\alpha(x,x') = 1 \wedge \frac{\pi(x')}{\pi(x)}.
\end{equation}
In addition to simplifying calculations, (\ref{eqn:RWM2}) strengthens the intuition for the method, since proposed moves with higher density under $\pi(\cdot)$ will always be accepted.  A typical choice for $Q(x,\cdot)$ is $\mathcal{N}(x,\lambda^2\Sigma)$, where the matrix $\Sigma$ is often chosen in an attempt to match the correlation structure of $\pi(\cdot)$, or simply taken as the identity \cite{sherlock2010random}.  The tuning parameter $\lambda$ is the only other user-specific input required.

\begin{figure}[ht]
\centering
\includegraphics[width=13.5cm, height=6.5cm]{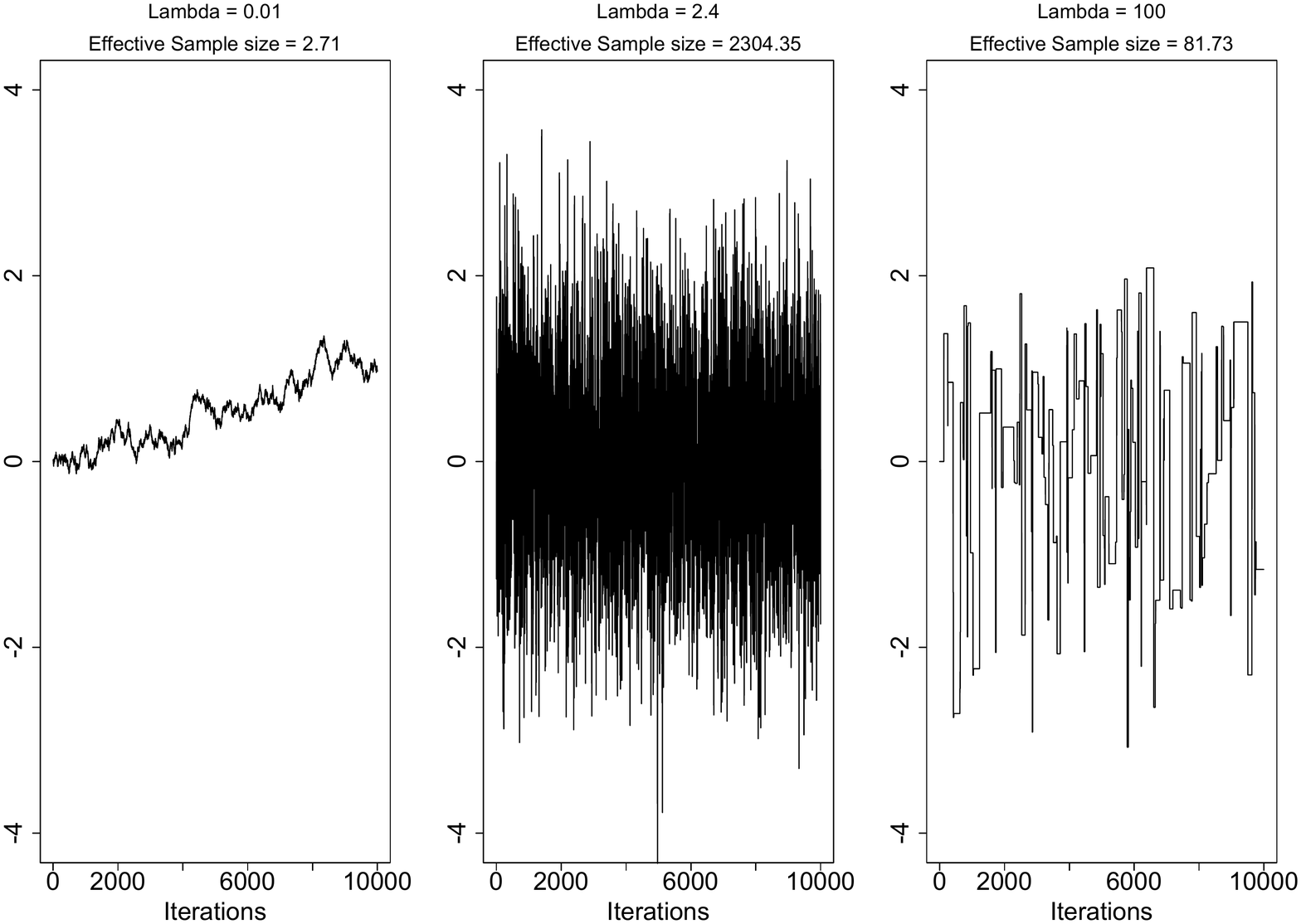}
\caption{ \normalsize These \emph{traceplots} show the evolution of three RWM Markov chains for which $\pi(\cdot)$ is a $\mathcal{N}(0,1^2)$ distribution, with different choices for $\lambda$.}
\label{fig:1DN}
\end{figure}

Much research has been conducted into properties of the \emph{random walk Metropolis} algorithm (RWM).  It has been shown that the optimal acceptance rate for proposals tends to $0.234$ as the dimension $n$ of the state space $\mathcal{X}$ tends to $\infty$ for a wide class of targets (e.g. \cite{sherlock2009optimal, sherlock2013optimal}).  The intuition for an optimal acceptance rate is to find the right balance between the distance of proposed moves and the chances of acceptance.  Increasing the former will reduce the autocorrelation in the chain if the proposal is accepted, but if it is rejected the chain will not move at all, so autocorrelation will be high.  Random walk proposals are sometimes referred to as \emph{blind} (e.g. \cite{beskos2013advanced}), as no information about $\pi(\cdot)$ is used when generating proposals, so typically very large moves will result in a very low chance of acceptance, while small moves will be accepted but result in very high autocorrelation for the chain.  Figure \ref{fig:1DN} demonstrates this in the simple case where $\pi(\cdot)$ is a one dimensional $\mathcal{N}(0,1^2)$ distribution.

Several authors have also shown that for certain classes of $\pi(\cdot)$ the tuning parameter $\lambda$ should be chosen such that $\lambda^2 \propto n^{-1}$ so that $\alpha \nrightarrow 0$ as $n \rightarrow \infty$ \cite{roberts2001optimal}. Because of this we say that algorithm efficiency `scales' $O(n^{-1})$ as the dimension $n$ of $\pi(\cdot)$ increases.  

Ergodicity results for a Markov chain constructed using the RWM algorithm also exist \cite{roberts1996geometric, mengersen1996rates, jarner2000geometric}.  At least exponentially light tails are a necessity for $\pi(x)$ for geometric ergodicity, which means that $\pi(x)/e^{-\|x\|} \to c$ as $\|x\| \to \infty$, for some constant $c$.  For super-exponential tails (where $\pi(x) \to 0$ at a faster than exponential rate), additional conditions are required \cite{roberts1996geometric, jarner2000geometric}.  We demonstrate with a simple example why heavy-tailed forms of $\pi(x)$ pose difficulties here (where $\pi(x) \to 0$ at a rate \emph{slower} then $e^{-\|x\|}$).

\vspace{0.5cm}

{\itshape \noindent \textbf{Example:} Take $\pi(x) \propto 1/(1+x^2)$, so that $\pi(\cdot)$ is a Cauchy distribution.  Then  if $X' \sim \mathcal{N}(x,\lambda^2)$, the ratio $\pi(x')/\pi(x) = (1 + (x')^2)/(1+ x^2) \to 1$ as $|x| \to \infty$.  So if $x_0$ is far away from $0$, the Markov chain will dissolve into a random walk, with almost every proposal being accepted.}

\vspace{0.5cm}

In the remainder of the article we will primarily discuss another approach to choosing $Q$ which has been shown empirically \cite{girolami2011riemann} and in some cases theoretically \cite{roberts2001optimal} to be superior to the RWM algorithm, though it should be noted that random walk proposals are still widely used in practice and are often sufficient for more straightforward problems \cite{sherlock2010random}.

\section{Diffusions} \label{sec:diffusions}

In MCMC we are concerned with discrete time processes.  However, often there are benefits to first considering a continuous time process with properties we desire.  One such is that some continuous time processes can be specified via a form of differential equation.  In this section we derive a choice for a Metropolis--Hastings proposal kernel based on approximations to \emph{diffusions}, those continuous-time $n$-dimensional Markov processes $(X_t)_{t \geq 0}$ for which any sample path $t \mapsto X_t(\omega)$ is a continuous function with probability one.  For any fixed $t$, we assume $X_t$ is a random variable taking values on the measurable space $(\mathcal{X},\mathcal{B})$ as before.  In the next section we provide some preliminaries, followed by an introduction to our main object of study, the Langevin diffusion.

\subsection{Preliminaries}

We focus on the class of time-homogeneous It\^{o} diffusions, whose dynamics are governed by a stochastic differential equation of the form
\begin{equation} \label{eqn:SDE}
dX_t = b(X_t)dt + \sigma(X_t)dB_t, ~~ X_0 = x_0,
\end{equation}
where $(B_t)_{t \geq 0}$ is a standard Brownian motion and the \emph{drift} vector $b$ and \emph{volatility} matrix $\sigma$ are Lipschitz continuous \cite{oksendal2003stochastic}.  Since $\mathbb{E}[B_{t+ \triangle t} - B_t|B_t = b_t] = 0$ for any  $\triangle t \geq 0$, informally we can see that
\begin{equation}
\mathbb{E}[X_{t + \triangle t} - X_t|X_t = x_t] = b(x_t)\triangle t + o(\triangle t),
\end{equation}
implying that the drift dictates how the mean of the process changes over a small time interval, and if we define the process $(M_t)_{t \geq 0}$ through the relation
\begin{equation}
M_t = X_t - \int_0^t b(X_s)ds
\end{equation}
then we have
\begin{equation}
\mathbb{E}[(M_{t + \triangle t} - M_t)(M_{t + \triangle t} - M_t)^T|M_t = m_t, X_t=x_t] = \sigma(x_t)\sigma(x_t)^T \triangle t + o(\triangle t),
\end{equation}
giving the \emph{stochastic} part of the relationship between $X_{t + \triangle t}$ and $X_t$ for small enough $\triangle t$, see e.g. \cite{rogers2000diffusions}.

While (\ref{eqn:SDE}) is often a suitable description of an It\^{o} diffusion, it can also be characterised through an \emph{infinitessimal generator} $\mathcal{A}$, which describes how functions of the process are expected to evolve.  We define this partial differential operator through its action on a function $f \in C_0(\mathcal{X})$ as
\begin{equation}
\mathcal{A}f(X_t) = \lim_{\triangle t \to 0} \frac{\mathbb{E}[f(X_{t + \triangle t})|X_t = x_t] - f(x_t)}{\triangle t},
\end{equation}
though $\mathcal{A}$ can be associated with the drift and volatility of $(X_t)_{t \geq 0}$ by the relation
\begin{equation} \label{eqn:gen}
\mathcal{A}f(x) = \sum_i b_i(x)\pd{f}{x_i}(x) + \frac{1}{2}\sum_{i,j} V_{ij}(x)\pd{^2 f}{x_i \partial x_j}(x),
\end{equation}
where $V_{ij}(x)$ denotes the component in row $i$ and column $j$ of $\sigma(x)\sigma(x)^T$ \cite{oksendal2003stochastic}.  Later on we shall use the generator characterisation of a diffusion to generalise it in some sense.

As in the discrete case, we can describe the transition kernel of a continuous time Markov process $P^t(x_0,\cdot)$.  In the case of an It\^{o} diffusion, $P^t(x_0,\cdot)$ admits a density $p_t(x|x_0)$, which in fact varies smoothly as a function of $t$.  The \emph{Fokker--Planck} equation describes this variation in terms of the drift and volatility, and is given by
\begin{equation} \label{eqn:FPK}
\pd{}{t}p_t(x|x_0) = - \sum_i \pd{}{x_i}[b_i(x)p_t(x|x_0)] + \frac{1}{2}\sum_{i,j} \pd{^2}{x_i \partial x_j}[V_{ij}(x)p_t(x|x_0)].
\end{equation}
Although typically the form of $P^t(x_0,\cdot)$ is unknown the expectation and variance of $X_t \sim P^t(x_0,\cdot)$ are given by the integral equations:
\begin{align*}
\mathbb{E}[X_t|X_0 = x_0] &= x_0 + \mathbb{E}\left[ \int_0^t b(X_s)ds \right], \\
\mathbb{E}[ (X_t - \mathbb{E}[X_t])(X_t - \mathbb{E}[X_t])^T|X_0 = x_0] &= \mathbb{E}\left[ \int_0^t \sigma(X_s)\sigma(X_s)^T ds \right],
\end{align*}
where the second of these is a result of the It\^{o} isometry \cite{oksendal2003stochastic}.  Continuing the analogy, a natural question is whether a diffusion process has an invariant distribution $\pi(\cdot)$, and whether
\begin{equation}
\lim_{t \to \infty} P^t(x_0,A) = \pi(A)
\end{equation}
for any $A \in \mathcal{B}$ and any $x_0 \in \mathcal{X}$, in some sense.  For a large class of diffusions (which we confine ourselves to) this is in fact the case \cite{meyn1993stability}, and in addition (\ref{eqn:FPK}) provides a means of finding $\pi(\cdot)$ given $b$ and $\sigma$.  Setting the left-hand side of (\ref{eqn:FPK}) to zero gives
\begin{equation} \label{eqn:FPK2}
\sum_i \pd{}{x_i}[b_i(x)\pi(x)] = \frac{1}{2}\sum_{i,j} \pd{^2}{x_i \partial x_j}[V_{ij}(x)\pi(x)],
\end{equation}
which can be solved to find $\pi(\cdot)$.

\subsection{Langevin diffusions}

Given (\ref{eqn:FPK2}) our goal becomes clearer: find drift and volatility terms so that the resulting dynamics describe a diffusion which converges to some user-defined invariant distribution $\pi(\cdot)$.  This process can then be used as a basis for choosing $Q$ in a Metropolis--Hastings algorithm.  The Langevin diffusion, first used to describe the dynamics of molecular systems \cite{coffey2004langevin}, is such a process, given by the solution to the stochastic differential equation
\begin{equation} \label{eqn:lang}
dX_t = \frac{1}{2}\nabla \log\pi(X_t)dt + dB_t, ~~ X_0 = x_0.
\end{equation}
Since $V_{ij}(x) = \mathbbm{1}_{\{i=j\}}$, it is clear that
\begin{equation}
\frac{1}{2}\pd{}{x_i}[\log\pi(x)]\pi(x) = \frac{1}{2}\pd{}{x_i}\pi(x), ~~ \forall i,
\end{equation}
which is a sufficient condition for (\ref{eqn:FPK2}) to hold.  So for any case in which $\pi(x)$ is suitably regular so that $\nabla\log\pi(x)$ is well-defined and the derivatives in (\ref{eqn:FPK2}) exist, we can use (\ref{eqn:lang}) to construct a diffusion which has invariant distribution $\pi(\cdot)$.

Roberts \& Tweedie \cite{roberts1996exponential} give sufficient conditions on $\pi(\cdot)$ under which a diffusion $(X_t)_{t \geq 0}$ with dynamics given by (\ref{eqn:lang}) will be ergodic, meaning
\begin{equation}
\| P^t(x_0,\cdot) - \pi(\cdot) \|_{TV} \to 0
\end{equation}
as $t \to \infty$, for any $x_0 \in \mathcal{X}$.  They remark that these conditions `should be appropriate for virtually all commonly encountered target densities' \cite{roberts1996exponential}.

\subsection{Metropolis-adjusted Langevin algorithm}

We can use Langevin diffusions as a basis for MCMC in many ways, but a popular variant is known as the Metropolis-adjusted Langevin algorithm (MALA), whereby $Q(x,\cdot)$ is constructed through an Euler--Maruyama discretisation of (\ref{eqn:lang}) and used as a candidate kernel in a Metropolis--Hastings algorithm.  The resulting $Q$ is
\begin{equation} \label{eqn:langprop}
Q(x,\cdot) \equiv \mathcal{N}\left(x + \frac{\lambda^2}{2}\nabla\log\pi(x), \lambda^2 I \right),
\end{equation}
where $\lambda$ is again a tuning parameter.

Before we discuss the theoretical properties of the approach, we first offer intuition for the dynamics.  From (\ref{eqn:langprop}) it can be seen that Langevin-type proposals comprise a deterministic shift towards a local mode of $\pi(x)$, combined with some random additive Gaussian noise, with variance $\lambda^2$ for each component.  The relative weights of the deterministic and random parts are fixed, given as they are by the parameter $\lambda$.  Typically if $\lambda^{1/2} \gg \lambda$ then the random part of the proposal will dominate, and vice versa in the opposite case, though this also depends on the form of $\nabla\log\pi(x)$ \cite{roberts1996exponential}. 

Again since this is a Metropolis--Hastings method, choosing $\lambda$ is a balance between proposing large enough jumps and ensuring that a reasonable proportion are accepted.  It has been shown that in the limit as $n \to \infty$ the optimal acceptance rate for the algorithm is $0.574$ \cite{roberts2001optimal} for forms of $\pi(\cdot)$ which either have independent and identically distributed components or whose components only differ by some scaling factor \cite{roberts2001optimal}.  In these cases, as $n \to \infty$ the parameter $\lambda$ must be $\propto n^{-1/3}$, so we say algorithm efficiency scales $O(n^{-1/3})$.  Note that these results compare favourably with the $O(n^{-1})$ scaling of the random walk algorithm.

Convergence properties of the method have also been established.  Roberts \& Tweedie \cite{roberts1996exponential} highlight some cases in which MALA is either geometrically ergodic or not.  Typically results are based on the tail behaviour of $\pi(x)$.  If these tails are heavier than exponential, then the method is typically \emph{not} geometrically ergodic, and similarly if the tails are lighter than Gaussian.  However, in the in between case the converse is true.  We again offer two simple examples for intuition here.

\vspace{0.5cm}

{\itshape \noindent \textbf{Example:} Take $\pi(x) \propto 1/(1+x^2)$ as in the previous example.  Then $\nabla\log\pi(x) = -2x/(1+x^2)^2 \to 0$ as $|x| \to \infty$.  So if $x_0$ is far away from 0, then the MALA will be approximately equal to the RWM algorithm, and so will also dissolve into a random walk.} 

\vspace{0.5cm}

{\itshape \noindent \textbf{Example:} Take $\pi(x) \propto e^{-x^4}$.  Then $\nabla\log\pi(x) = -4x^3$ and $X' \sim \mathcal{N}(x - 4\lambda^2x^3, \lambda^2)$.  So for any fixed $\lambda$, there exists $c >0$ such that for $|x_0| > c$ we have $|4\lambda^2x^3| >> x$ and $|x - 4\lambda^2x^3| >> \lambda$, suggesting that MALA proposals will quickly spiral further and further away from any neighbourhood of $0$, and hence nearly all will be rejected.}

\vspace{0.5cm}

For cases where there is strong correlation between elements of $x$ or each element has a different marginal variance, the MALA can also be `pre-conditioned' in a similar way to the RWM, so that the covariance structure of proposals more accurately reflects that of $\pi(x)$ \cite{roberts2002langevin}.  In this case, proposals take the form
\begin{equation} \label{eqn:langprop2}
Q(x,\cdot) \equiv \mathcal{N}\left(x + \frac{\lambda^2}{2}\Sigma\nabla\log\pi(x), \lambda^2 \Sigma \right),
\end{equation}
where $\lambda$ is again a tuning parameter.  It can be shown that provided $\Sigma$ is a constant matrix, $\pi(x)$ is still the invariant distribution for the diffusion on which (\ref{eqn:langprop2}) is based \cite{xifara2013langevin}.

\section{Geometric concepts in Markov Chain Monte Carlo} \label{sec:geometry}

Ideas from Information geometry have been successfully applied to Statistics from as early as \cite{jeffreys1946invariant}.  More widely, other geometric ideas have also been applied, offering new insight into common problems (e.g. \cite{critchley1993preferred, marriott2002local}).  A survey is given in \cite{barndorff1986role}.  In this section we suggest why some ideas from differential geometry may be beneficial for sampling methods based on Markov chains.  We then review what is meant by a `diffusion on a manifold', before turning to the specific case of (\ref{eqn:lang}).  After this we discuss what can be learned from work in Information geometry in this context.

\subsection{Manifolds and Markov chains}

We often make assumptions in MCMC about the properties of the space $\mathcal{X}$ in which our Markov chains evolve.  Often $\mathcal{X} = \mathbb{R}^n$, and in many cases where it is not then a simple re-parametrisation would make it so.  But here by $\mathbb{R}^n$ we mean the set of ordered $n$-tuples of real numbers, $\mathbb{R}^n = \{ (a_1,...,a_n) : a_i \in (-\infty,\infty) ~ \forall i \}$.  The additional assumption that is often made is that $\mathbb{R}^n$ is \emph{Euclidean}, an inner product space with the induced distance metric
\begin{equation} \label{eqn:emetric}
d(x,y) = \sqrt{ \sum_i (x_i - y_i)^2 }.
\end{equation}
For sampling methods based on Markov chains which explore the space \emph{locally}, like the RWM and MALA, it may be advantageous to instead impose a different metric structure on the space $\mathcal{X}$, so that some points are drawn closer together and others pushed further apart.  Intuitively, one can picture distances in the space being defined such that if the current position in the chain is far from an area of $\mathcal{X}$ which is `likely to occur' under $\pi(\cdot)$, then the distance to such a \emph{typical set} could be reduced.  Similarly, once this region is reached, the space could be `stretched` or `warped' so that it is explored as efficiently as possible.

While the idea is attractive, it is far from a constructive definition.  We only have the pre-requisite that $(\mathcal{X},d)$ must be a metric space.  However, as Langevin dynamics use \emph{gradient} information, we will require $(\mathcal{X},d)$ to be a space on which we can do differential calculus.  Riemannian manifolds are an appropriate choice therefore, as the rules of differentiation are well understand for functions defined on them \cite{boothby1986introduction, lee2003smooth}, while we are still free to define a more \emph{local} notion of distance than Euclidean.  In this section we write $\mathbb{R}^n$ to denote the Euclidean vector space.

\subsection{Preliminaries}

We do not provide a full overview of Riemannian geometry here \cite{boothby1986introduction, lee2003smooth, do1992riemannian}.  We simply note that for our purposes we can consider an $n$-dimensional Riemannian manifold (henceforth manifold) to be an $n$-dimensional metric space, in which distances are defined in a specific way.  We also only consider manifolds for which a global coordinate chart exists, meaning that a mapping $r: \mathbb{R}^n \to M$ exists which is both differentiable and invertible, and for which the inverse is also differentiable (a \emph{diffeomorphism}).  Although this restricts the class of manifolds available (the sphere, for example, is not in this class), it is again suitable for our needs, and avoids the practical challenges of switching between coordinate patches.  The connection with $\mathbb{R}^n$ defined through $r$ is crucial for making sense of differentiability in $M$.  We say a function $f : M \to \mathbb{R}$ is `differentiable' if $(f \circ r): \mathbb{R}^n \to \mathbb{R}$ is \cite{lee2003smooth}.

As has been stated, (\ref{eqn:emetric}) can be induced via a Euclidean inner product, which we denote $\langle \cdot,\cdot \rangle$.  However, it will aid intuition to think of distances in $\mathbb{R}^n$ via curves 
\begin{equation}
\gamma: [0,1] \to \mathbb{R}^n.
\end{equation}
We \emph{could} think of the distance between two points in $x,y \in \mathbb{R}^n$ as the minimum length among all curves which pass through $x$ and $y$.  If $\gamma(0) = x$ and $\gamma(1) = y$ the length is defined as
\begin{equation} \label{eqn:length}
L(\gamma) = \int_0^1 \sqrt{ \langle \gamma'(t), \gamma'(t) \rangle } dt,
\end{equation}
giving the metric
\begin{equation} \label{eqn:rmetric}
d(x,y) = \inf \left\{ L(\gamma) : \gamma(0) = x, \gamma(1) = y \right\}.
\end{equation}
In $\mathbb{R}^n$ the curve with minimum length will be a straight line, so that (\ref{eqn:rmetric}) agrees with (\ref{eqn:emetric}).  More generally we call a solution to (\ref{eqn:rmetric}) a geodesic \cite{boothby1986introduction}.

In a vector space, metric properties can always be induced through an inner product (which also gives a notion of orthogonality).  Such a space can be thought of as `flat', since for any two points $y$ and $z$, the straight line $ay + (1-a)z, ~ a \in [0,1]$ is also contained in the space.  In general, manifolds do not have vector space structure globally, but do so at the infinitessimal level.  As such we can think of them as `curved'.  We cannot always define an inner product, but we can still define distances through (\ref{eqn:rmetric}).  We define a curve on a manifold $M$ as $\gamma_M: [0,1] \to M$.  At each point $\gamma_M(t) = p \in M$ the velocity vector $\gamma_M'(t)$ lies in an $n$-dimensional vector space which touches $M$ at $p$.  These are known as \emph{tangent spaces}, denoted $T_pM$,  which can be thought as local linear approximations to $M$.  We can define an inner product on each as $g_p: T_pM \to \mathbb{R}$, which allows us to define a generalisation of (\ref{eqn:length}) as
\begin{equation}
L(\gamma_M) = \int_0^1 \sqrt{g_p(\gamma_M'(t),\gamma_M'(t))} dt.
\end{equation}
and provides a means to define a metric on the manifold as $d(x,y) = \inf \left\{ L(\gamma_M) : \gamma_M(0) = x, \gamma_M(1) = y \right\}$.

\subsubsection{Embeddings and local coordinates}
Much of Riemannian geometry is written in a `coordinate free' manner, and \emph{coordinate invariance} is a pivotal concept in the subject \cite{lee2003smooth}.  For our purposes though, it suits calculations to work with a defined set of coordinates.  We use the term `local' coordinates here, since manifolds need only be locally mappable to $\mathbb{R}^n$.  Although we restrict study to manifolds which are \emph{globally} mappable to $\mathbb{R}^n$, we still use this convention.

So far we have introduced manifolds as abstract objects.  In fact, they can also be considered as objects which are \emph{embedded} in some higher-dimensional Euclidean space.  A simple example is any two-dimensional surface, such as the unit sphere, lying in $\mathbb{R}^3$.  If a manifold is embedded in this way, then metric properties can be induced from the ambient Euclidean space.  The Nash embedding theorem \cite{nash2002imbedding} states that any manifold can be embedded in such a way.

\begin{figure}[ht]
\centering
\includegraphics[trim=0cm 4cm 0cm 1.5cm, clip=true, width=13cm]{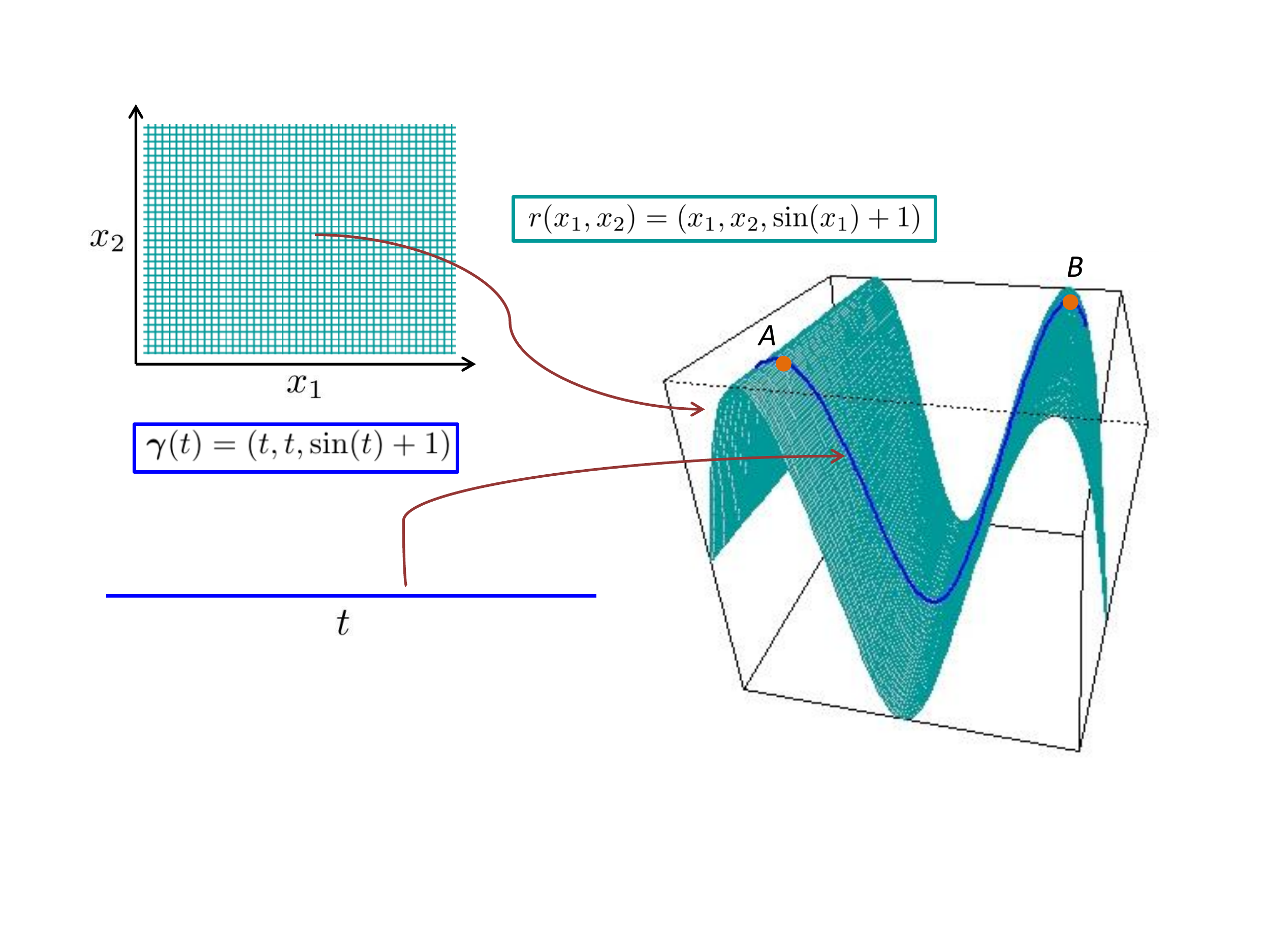}
\caption{ \normalsize A two-dimensional manifold (surface) embedded in $\mathbb{R}^3$ through $r(x_1,x_2) = (x_1,x_2,\sin(x_1)+1)$, parametrised by the local coordinates $x_1$ and $x_2$.  The distance between points $A$ and $B$ is given by the length of the curve $\gamma(t) = (t,t,\sin(t)+1))$.}
\end{figure}

We seek to make these ideas more concrete through an example, the graph of a function $f(x_1,x_2)$ of two variables $x_1$ and $x_2$.  The resulting map $r$ is
\begin{align}
r: \mathbb{R}^2 &\to M \\
r(x_1,x_2) &= (x_1,x_2,f(x_1,x_2)).
\end{align}
We can see that $M$ is embedded in $\mathbb{R}^3$, but that any point can be identified using only two coordinates $x_1$ and $x_2$.  In this case, each $T_pM$ is a plane, and therefore a two-dimensional subspace of $\mathbb{R}^3$, so (i) it inherits the Euclidean inner product $\langle \cdot, \cdot \rangle$ and (ii) any vector $v \in T_pM$ can be expressed as a linear combination of any two linearly independent \emph{basis} vectors: a canonical choice is the partial derivatives $\partial r/\partial x_1:=r_1$ and $r_2$ evaluated at $x = r^{-1}(p) \in \mathbb{R}^2$.  The resulting inner product $g_p(v,w)$ between two vectors $v,w \in T_pM$ can be induced from the Euclidean inner product, as
\begin{align*}
\langle v,w \rangle &= \langle v_1r_1(x) + v_2r_2(x), w_1r_1(x) + w_2r_2(x) \rangle, \\
&= v_1 w_1 \langle r_1(x),r_1(x) \rangle + v_1w_2\langle r_1(x),r_2(x) \rangle + v_2 w_1\langle r_2(x),r_1(x)\rangle + v_2 w_2 \langle r_2(x),r_2(x) \rangle, \\ \label{eqn:metric}
&= v^T G(x) w,
\end{align*}
where
\begin{equation}
G(x) = \left(
\begin{array}{cc} \langle r_1(x),r_1(x) \rangle & \langle r_1(x),r_2(x) \rangle \\ \langle r_1(x),r_2(x) \rangle & \langle r_2(x),r_2(x) \rangle \end{array}
\right)
\end{equation}
and we use $v_i, w_i$ to denote the components of $v$ and $w$.  To write (\ref{eqn:length}) using this notation, we define the curve $x(t) \in \mathbb{R}^2$ corresponding to $\gamma_M(t) \in M$ as $x = (r^{-1} \circ \gamma_M): [0,1] \to \mathbb{R}^2$.  Equation (\ref{eqn:length}) can then be written
\begin{equation}
L(\gamma_M) = \int_0^1 \sqrt{ x'(t)^T G(x(t)) x'(t) } dt,
\end{equation}
which can be used in (\ref{eqn:rmetric}) as before.

The key point is that although we have started with an object embedded in $\mathbb{R}^3$, we can compute the Riemannian metric $g_p(v,w)$ (and hence distances in $M$) using only the two-dimensional `local' coordinates $(x_1,x_2)$.  We also need not have explicit knowledge of the mapping $r$, only the components of the positive definite matrix $G(x)$.  The Nash embedding theorem in essence enables us to define manifolds by the reverse process: simply choose the matrix $G(x)$ so that we define a metric space with suitable distance properties, and some object embedded in some higher-dimensional Euclidean space will exist for which these metric properties can be induced as above.  So to define our new space, we simply choose an appropriate matrix-valued map $G(x)$ (we discuss this choice in Subsection \ref{sub:metric}).  If $G(x)$ does not depend on $x$, then $M$ has a vector space structure, and can be thought of as `flat'.  Trivially, $G(x) = I$ gives Euclidean $n$-space.

We can also define volumes on a Riemannian manifold in local coordinates.  Following standard coordinate transformation rules, we can see that for the above example the area element $dx$ in $\mathbb{R}^2$ will change according to a jacobian $J = |(Dr)^T(Dr)|^{1/2}$, where $Dr = \partial(p_1,p_2,p_3)/\partial(x_1,x_2)$.  This reduces to $J = |G(x)|^{1/2}$, which is also the case for more general manifolds \cite{boothby1986introduction}.  We therefore define the Riemannian volume measure on a manifold $M$ in local coordinates as
\begin{equation}
\text{Vol}_M(dx) = |G(x)|^{\frac{1}{2}}dx.
\end{equation}
If $G(x) = I$ then this reduces to Lebesgue measure.

\subsection{Diffusions on manifolds}

By a `diffusion on a manifold' in local coordinates, we actually mean a diffusion on Euclidean space which, when mapped onto $M$ through $r$, becomes the desired diffusion \emph{along the manifold}.  For example, a `Brownian motion on a sphere' means the diffusion on Euclidean space that, when mapped onto the sphere, produces Brownian motion along the surface.  If a realisation of Brownian motion was drawn in wet ink on the surface of the sphere, and the sphere was then rolled over a piece of flat paper, the resulting imprint on the paper would be a realisation of Brownian motion on Euclidean space \cite{manton2013primer}.  However, the mapping $r$ does not correspond to this rolling procedure, so the pre-image of the Brownian motion on the sphere under $r$ will \emph{not} be Brownian motion on Euclidean space.

Our goal, therefore, is to define a diffusion on Euclidean space which, when mapped onto a manifold through $r$, becomes the Langevin diffusion described in (\ref{eqn:lang}) by the above procedure.  Such a diffusion takes the form
\begin{equation} \label{eqn:rlang}
dX_t = \frac{1}{2}\tilde{\nabla} \log \tilde{\pi}(X_t)dt + d\tilde{B}_t,
\end{equation}
where those objects marked with a tilde must be defined appropriately.  The next few paragraphs are technical, and readers aiming to simply grasp the key points may wish to skip to the end of this Subsection.

We turn first to $(\tilde{B}_t)_{t \geq 0}$, which we use to denote Brownian motion on a manifold.  Intuitively, we may think of a construction based on embedded manifolds, by setting $\tilde{B}_0 = p \in M$, and for each increment sampling some random vector in the tangent space $T_pM$, and then moving along the manifold in the prescribed direction for an infinitessimal period of time before re-sampling another velocity vector from the next tangent space \cite{manton2013primer}.  In fact we can define such a construction using Stratonovitch calculus, and show that the infinitessimal generator can be written using only local coordinates \cite{rogers2000diffusions}.  Here we instead take the approach of generalising the generator directly from Euclidean space to the local coordinates of a manifold, arriving at the same result.  We then deduce the stochastic differential equation describing $(\tilde{B}_t)_{t \geq 0}$ in It\^{o} form using (\ref{eqn:gen}).

For a standard Brownian motion on $\mathbb{R}^n$, $\mathcal{A} = \triangle /2$, where $\triangle$ denotes the Laplace operator
\begin{equation} \label{eqn:laplacian}
\triangle f = \sum_i \pd{^2 f}{x_i^2} = \text{div}( \nabla f).
\end{equation}
Substituting $\mathcal{A}=\triangle /2$ into (\ref{eqn:gen}) trivially gives $b_i(x) = 0 \: \: \forall i$, $V_{ij}(x) = \mathbbm{1}_{\{i=j\}}$ as required.  The Laplacian $\triangle f(x)$ is the divergence of the gradient vector field of some function $f \in C^2(\mathbb{R}^n)$, and its value at $x \in \mathbb{R}^n$ can be thought of as the average value of $f$ in some neighbourhood of $x$ \cite{stewart2011multivariable}.

To define a Brownian motion on any manifold, the gradient and divergence operators must be generalised.  The \emph{gradient} of a function on $\mathbb{R}^n$ is the unique vector field such that for any unit vector $u$
\begin{equation} \label{eqn:direc}
\langle \nabla f(x),u \rangle = D_{u}\left[f(x)\right] = \lim_{h \to 0} \left\{ \frac{f(x+hu) - f(x)}{h} \right\},
\end{equation}
the directional derivative of $f$ along $u$ at $x \in \mathbb{R}^n$.

On a manifold, the gradient operator $\nabla_M$ can still be defined such that the inner product $g_p(\nabla_M f(x),u ) = D_u [f(x)]$. Setting $\nabla_M = G(x)^{-1}\nabla$ gives
\begin{align*}
g_p(\nabla_M f(x),u) &= (G^{-1}(x)\nabla f(x))^T G(x) u, \\
&= \langle \nabla f(x), u \rangle,
\end{align*}
which is equal to the directional derivative along $u$ as required.

The \emph{divergence} of some vector field $v$ at a point $x \in \mathbb{R}^n$ is the net outward flow generated by $v$ through some small neighbourhood of $x$.  Mathematically, the divergence of $v(x) \in \mathbb{R}^3$ is given by $\sum_i \partial v_i/\partial x_i$.  On a more general manifold the divergence is also a sum of derivatives, but here they are \emph{covariant} derivatives.  A short introduction is provided in Appendix \ref{app:A}.  Here we simply state that the covariant derivative of a vector field $v$ at a point $p \in M$ is the orthogonal projection of the directional derivative onto the tangent space $T_pM$.  Intuitively, a vector field on a manifold is a field of vectors each of which lie in the tangent space to a point $p \in M$.  It only makes sense therefore to discuss how vector fields change \emph{along} the manifold, or in the direction of vectors which also lie in the tangent space.  Although the idea seems simple, the covariant derivative has some attractive geometric properties, notably it can be completely written in local coordinates, and so does not depend on knowledge of an embedding in some ambient space.

The divergence of a vector field $v$ defined on a manifold $M$ at the point $p \in M$ is defined as:
\begin{equation} \label{eqn:div2}
\text{div}_M(v) = \sum_{i=1}^n D^c_{e_i} [v_i],
\end{equation}
where $e_i$ denotes the $i$th basis vector for the tangent space $T_pM$ at $p \in M$, and $v_i$ denotes the $i$th coefficient.  This can be written in local coordinates (see Appendix \ref{app:A}) as
\begin{equation} \label{eqn:divRM}
\text{div}_M(v) = |G(x)|^{-\frac{1}{2}} \sum_{i=1}^n \frac{\partial}{\partial x_i} \left( |G(x)|^{\frac{1}{2}} v_i \right).
\end{equation}
Combining these two operators, we can define a generalisation of the Laplace operator, known as the \emph{Laplace--Beltrami} operator (e.g. \cite{hsu2002stochastic, kent1978time}), as:
\begin{equation}
\triangle_{LB}f = \text{div}_M(\nabla_M f) = |G(x|^{-\frac{1}{2}} \sum_{i=1}^n \frac{\partial}{\partial x_i} \left( |G(x)|^{\frac{1}{2}} \sum_{j=1}^n \{G^{-1}(x)\}_{ij} \frac{\partial f}{\partial x_j} \right),
\end{equation}
for some $f \in C_0^2(M)$.

The generator of a Brownian motion on $M$ is $\triangle_{LB}/2$ \cite{hsu2002stochastic}.  Using (\ref{eqn:gen}) the resulting diffusion has dynamics given by
\begin{align*}
d\tilde{B}_t &= \Omega(X_t)dt + \sqrt{G^{-1}(X_t)}dB_t, \\
\Omega_i(X_t) &= \frac{1}{2}|G(X_t)|^{-\frac{1}{2}} \sum_{j=1}^n \frac{\partial}{\partial x_j} \left( |G(X_t)|^{\frac{1}{2}}\{G^{-1}(X_t)\}_{ij} \right).
\end{align*}
Those familiar with the It\^{o} formula will not be surprised by the additional drift term $\Omega(X_t)$.  As It\^{o} integrals do not follow the chain rule of ordinary calculus, non-linear mappings of martingales like $(B_t)_{t \geq 0}$ typically result in drift terms being added to the dynamics (e.g. \cite{oksendal2003stochastic}).

To define $\tilde{\nabla}$ we simply note that this is again the gradient operator on a general manifold, so $\tilde{\nabla} = G^{-1}(x)\nabla$.  For the density $\tilde{\pi}(x)$, we note that this density will now implicitly be defined with respect to the volume measure on the manifold.  So to ensure the diffusion (\ref{eqn:rlang}) has the correct invariant density with respect to Lebesgue measure, we define
\begin{equation}
\tilde{\pi}(x) = \pi(x)|G(x)|^{-\frac{1}{2}}.
\end{equation}
Putting these three elements together, (\ref{eqn:rlang}) becomes
\begin{equation*}
dX_t = \frac{1}{2}G^{-1}(X_t)\nabla\log\left\{\pi(X_t)|G(X_t)|^{-\frac{1}{2}}\right\}dt + \Omega(X_t)dt + \sqrt{G^{-1}(X_t)}dB_t,
\end{equation*}
which, upon simplification, becomes
\begin{align} \label{eqn:rlang2}
dX_t &= \frac{1}{2}G^{-1}(X_t)\nabla\log\pi(X_t)dt + \Lambda(X_t)dt + \sqrt{G^{-1}(X_t)}dB_t, \\
\Lambda_i(X_t) &= \frac{1}{2}\sum_j \pd{}{x_j} \{G^{-1}(X_t)\}_{ij}.
\end{align}
It can be shown that this diffusion has invariant Lebesgue density $\pi(x)$ as required \cite{xifara2013langevin}.  Intuitively when a set is mapped onto the manifold, distances are changed by a factor $\sqrt{G(x)}$.  So to end up with the initial distances, they must first be changed by a factor of $\sqrt{G^{-1}(x)}$ before the mapping, which explains the volatility term in (\ref{eqn:rlang2}).

The resulting Metropolis--Hastings proposal kernel for this `MALA on a manifold' was clarified in \cite{xifara2013langevin}, and is given by
\begin{equation} \label{eqn:mmala}
Q(x,\cdot) \equiv \mathcal{N}\left( x + \frac{\lambda^2}{2}G^{-1}(x)\nabla\log\pi(x) + \lambda^2 \Lambda(x), \lambda^2G^{-1}(x) \right),
\end{equation}
where $\lambda^2$ is a tuning parameter.  The nonlinear drift term here is slightly different to that reported in \cite{girolami2011riemann, roberts2002langevin}, for reasons discussed in \cite{xifara2013langevin}.

\subsection{Choosing a Metric} \label{sub:metric}

We now turn to the question of which manifold to choose, or equivalently how to choose $G(x)$.  In this section we sometimes switch notation slightly, denoting the target density $\pi(x|y)$, as some of the discussion is directed towards Bayesian inference, where $\pi(\cdot)$ is the posterior distribution for some parameter $x$ after observing some data $y$.  The problem statement is: what is an appropriate choice of distance between points in the sample space of a given probability distribution?  

A related (but distinct) question is how to define a distance between two probability distributions from the same parametric family but with different parameters.  This has been a key theme in Information geometry, explored by Rao \cite{radhakrishna1945information} and others \cite{amari2007methods} for many years.  Although generic measures of distance between distributions (such as total variation) are often appropriate, based on Information-theoretic principles one can deduce that for a given parametric family $\{ p_{x}(y) : x \in \mathcal{X} \}$, it is in some sense natural to consider this `space of distributions' to be a manifold, where the Fisher information is the matrix $G(x)$ (with the $\alpha = 0$ connection employed, see \cite{amari2007methods} for details).

Because of this, Girolami \& Calderhead \cite{girolami2011riemann} proposed a variant of the Fisher \emph{metric} for geometric Markov chain Monte Carlo, as
\begin{equation} \label{eqn:fisher}
G(x) = \mathbb{E}_{y|x} \left[ -\pd{^2}{x_i \partial x_j} \log f(y|x) \right] - \pd{^2}{x_i \partial x_j} \log\pi_0(x),
\end{equation}
where $\pi(x|y) \propto f(y|x)\pi_0(x)$ is the target density, $f$ denotes the likelihood and $\pi_0$ the prior.  The metric is tailored to Bayesian problems, which are a common use for MCMC, so the Fisher information is combined with the negative Hessian of the log-prior.  One can also view this metric as the expected negative Hessian of the log target, since this naturally reduces to (\ref{eqn:fisher}).

The motivation for a Hessian-style metric can also be understood from studying MCMC proposals.  From (\ref{eqn:mmala}) and by the same logic as for general pre-conditioning methods \cite{roberts2002langevin}, the objective is to choose $G^{-1}(x)$ to match the covariance structure of $\pi(x|y)$ locally.  If the target density were Gaussian with covariance matrix $\Sigma$, then
\begin{equation} \label{eqn:hess}
 - \pd{^2}{x_i \partial x_j} \log\pi(x|y) = \Sigma.
\end{equation}
In the non-Gaussian case, the negative Hessian is no longer constant, but we can imagine that it matches the correlation structure of $\pi(x|y)$ \emph{locally} at least.  Such ideas have been discussed in the geostatistics literature previously \cite{christensen2006robust}.  One problem with simply using (\ref{eqn:hess}) to define a metric is that unless $\pi(x|y)$ is log-concave the negative Hessian will not be globally positive-definite, although Petra et al. \cite{petra2013computational} conjecture that it may be appropriate for use in some realistic scenarios, and suggest some computationally efficient approximation procedures \cite{petra2013computational}.

\vspace{0.5cm}

{\itshape \noindent \textbf{Example:} Take $\pi(x) \propto 1/(1+x^2)$, and set $G(x) = -\partial^2\log\pi(x)/\partial x^2$.  Then $G^{-1}(x) = (1+x^2)^2/(2-2x^2)$, which is negative if $x^2 > 1$, so unusable as a proposal variance.}

\vspace{0.5cm}

Girolami \& Calderhead \cite{girolami2011riemann} use the Fisher metric in part to counteract this problem.  Taking expectations over the data ensures that the likelihood contribution to $G(x)$ in (\ref{eqn:fisher}) will be positive (semi-)definite globally (e.g. \cite{pawitan2001all}), so provided a log-concave prior is chosen then (\ref{eqn:fisher}) should be a suitable choice for $G(x)$.  Indeed, Girolami \& Calderhead \cite{girolami2011riemann} provide several examples in which geometric MCMC methods using this Fisher metric perform better than their `non-geometric' counterparts.

Betancourt \cite{betancourt2013general} also starts from the viewpoint that the Hessian (\ref{eqn:hess}) is an appropriate choice for $G(x)$, and defines a mapping from the set of $n \times n$ matrices to the set of positive-definite $n \times n$ matrices by taking a `smooth' absolute value of the eigenvalues of the Hessian.  This is done in a way such that derivatives of $G(x)$ are still computable, inspiring the author to the name \emph{SoftAbs} metric.  For a fixed value of $x$, the negative Hessian $H(x)$ is first computed, and then decomposed into $U^TDU$, where $D$ is the diagonal matrix of eigenvalues.  Each diagonal element of $D$ is then altered by the mapping $t_{\alpha}: \mathbb{R} \to \mathbb{R}$, given by
\begin{equation}
t_{\alpha}(\lambda_i) = \lambda_i \coth(\alpha \lambda_i),
\end{equation}
where $\alpha$ is a tuning parameter (typically chosen to be as large as possible for which eigenvalues remain non-zero numerically).  The function $t_{\alpha}$ acts as an absolute value function, but also uplifts eigenvalues which are close to zero to $\approx 1/\alpha$.  It should be noted that while the Fisher metric is only defined for models in which a likelihood is present, and for which the expectation is tractable, the SoftAbs metric can be found for any target distribution $\pi(\cdot)$.

Many authors (e.g. \cite{girolami2011riemann, petra2013computational}) have noted that for many problems the terms involving derivatives of $G(x)$ are often small, and so it is not always worth the computational effort of evaluating them.  Girolami \& Calderhead \cite{girolami2011riemann} propose the simplified manifold MALA, in which proposals are of the form
\begin{equation}
Q(x,\cdot) \equiv \mathcal{N}\left( x + \frac{\lambda^2}{2}G^{-1}(x) \nabla\log\pi(x), \lambda^2 G^{-1}(x) \right)
\end{equation}
Using this method means derivatives of $G(x)$ are no longer needed, so more pragmatic ways of regularising the Hessian are possible.  One simple approach would be to take the absolute values of each eigenvalue, giving $G(x) = U^T |D| U$, where $H(x) = U^TDU$ is the negative Hessian and $|D|$ is a diagonal matrix with $\{|D|\}_{ii} = |\lambda_i|$ (this approach may fall into difficulties if eigenvalues are numerically zero).  Another would be choose $G(x)$ as the `nearest' positive-definite matrix to the negative Hessian, according to some distance metric on the set of $n \times n$ matrices. The problem has in fact been well-studied in mathematical finance, in the context of finding correlations using incomplete data sets \cite{higham2002computing}, and tackled using distances induced by the Frobenius norm.  Approximate solution algorithms are discussed in Higham \cite{higham2002computing}.  We again provide a simple example suggesting that a `Hessian-style metric' can alleviate some of the difficulties associated with heavy-tailed target densities.

\vspace{0.5cm}

{\itshape \noindent \textbf{Example:} Take $\pi(x) \propto 1/(1+x^2)$, and set $G(x) = |-\partial^2\log\pi(x)/\partial x^2$|.  Then $G^{-1}(x)\nabla\log\pi(x) = -x(1+x^2)/|1-x^2|$, which no longer tends to $0$ as $|x| \to \infty$, suggesting a manifold variant of MALA with a Hessian-style metric may avoid some of the pitfalls of the standard algorithm.  Note that the drift may become very large if $|x| \approx 1$, but since this event occurs with probability $0$ we do not see it as a major cause for concern.}

\vspace{0.5cm}

{\itshape \noindent \textbf{Example:} Take $\pi(x) \propto e^{-x^4}$, and set $G(x) = |-\partial^2\log\pi(x)/\partial x^2|$.  Then $G^{-1}(x)\nabla\log\pi(x) = -x/3$, which is $O(x)$, so alleviates the problem of spiralling proposals for light-tailed targets demonstrated by MALA in an earlier example.}

\vspace{0.5cm}

Other choices for $G(x)$ have been proposed which are not based on the Hessian.  These have the advantage that gradients need not be computed (either analytically or using computational methods).  Sejdinovic et al. \cite{sejdinovic2013kernel} propose a Metropolis--Hastings method which can be viewed as a geometric variant of the RWM, where the choice for $G(x)$ is based on mapping samples to an appropriate \emph{feature} space, and performing principal component analysis on the resulting features to choose a local covariance structure for proposals.

If we consider the RWM with Gaussian proposals to be an Euler--Maruyama discretisation of Brownian motion on a manifold, then proposals will take the form $Q(x,\cdot) \equiv \mathcal{N}(x + \lambda^2 \Omega(x), \lambda^2 G^{-1}(x))$.  If we assume (like in the simplified manifold MALA) that $\Omega(x) \approx 0$, then we have proposals centred at the current point in the Markov chain with a local covariance structure (the full Hastings acceptance rate must now be used as $q(x'|x) \neq q(x|x')$ in general).

As no gradient information is needed, the Sejdinovic et al. metric can be used in conjunction with the pseudo-marginal MCMC algorithm, so that $\pi(x|y)$ need not be known exactly.  Examples from the article demonstrate the power of the approach \cite{sejdinovic2013kernel}.

\section{Survey of applications} \label{sec:applications}

Rather than conduct our own simulation study, we instead highlight some cases in the literature where geometric MCMC methods have been used with success.

Martin et al. \cite{martin2012stochastic} consider Bayesian inference for a Statistical inverse problem, in which a surface explosion causes seismic waves to travel down into the ground (the \emph{subsurface medium}).  Often the properties of the subsurface vary with distance from ground level or because of obstacles in the medium, in which case a fraction of the waves will scatter off these \emph{boundaries} and be reflected back up to ground level at later times.  The observations here are the initial explosion and the waves which return to the surface, together with return times.  The challenge is to infer the properties of the subsurface medium from this data.  The authors construct a likelihood based on the wave equation for the data, and perform Bayesian inference using a variant of the manifold MALA.  Figures are provided showing the local correlations present in the posterior, and therefore highlighting the need for an algorithm which can navigate the high density region efficiently.  Several methods are compared in the paper, but the variant of MALA which incorporates a local correlation structure is shown to be the most efficient, particularly as the dimension of the problem increases \cite{martin2012stochastic}.

Calderhead \& Girolami \cite{calderhead2011statistical} dealt with two models for biological phenomena based on nonlinear dynamical systems.  A model of circadian control in the \emph{Arabidopsis thaliana} plant comprised a system of six nonlinear differential equations, with twenty two parameters to be inferred.  Another model for cell signalling consisted of a system of six nonlinear differential equations with eight parameters, with inference complicated by the fact that observations of the model are not recorded directly \cite{calderhead2011statistical}.  The resulting inference was performed using RWM, MALA and geometric methods, with the results highlighting the benefits of taking the latter approach.  The simplified variant of MALA on a manifold is reported to have produced the most efficient inferences overall, in terms of effective sample size per unit of computational time.

Stathopoulos \& Girolami \cite{stathopoulos2013markov} considered the problem of inferring parameters in Markov jump processes.  In the paper a \emph{linear noise approximation} is shown, which can make inference in such models more straightforward, enabling an approximate likelihood to be computed.  Models based on chemical reaction dynamics are considered; one such from chemical kinetics contained four unknown parameters, another from gene expression consisting of seven.  Inference was performed using the RWM, the simplified manifold MALA and Hamiltonian methods, with the MALA reported as most efficient according to the chosen diagnostics.  The authors note that the simplified manifold method is both conceptually simple and able to account for local correlations, making it an attractive choice for inference \cite{stathopoulos2013markov}.

Konukoglu et al. \cite{konukoglu2011efficient} designed a method for personalising a generic model for a physiological process to a specific patient, using clinical data.  The personalisation took the form of patient-specific parameter inference.  The authors highlight some of the difficulties of this task in general, including the complexity of the models and relative sparsity of the datasets, which often result in a parameter identifiability issue \cite{konukoglu2011efficient}.  The example discussed in the paper is the \emph{Eikonal-Diffusion} model describing electrical activity in cardiac tissue, which results in a likelihood for the data based on a nonlinear partial differential equation, combined with observation noise \cite{konukoglu2011efficient}.  A method for inference was developed by first approximating the likelihood using a spectral representation, and then using geometric MCMC methods on the resulting approximate posterior.  The method was first evaluated on synthetic data, and the authors note that inference took `less than five minutes' to provide accurate estimates.  The method was then repeated on clinical data taken from a study for ventricular tachycardia radio-frequency ablation \cite{konukoglu2011efficient}.

\section{Discussion} \label{sec:discussion}

The geometric viewpoint in not necessary to understand manifold variants of the MALA.  Indeed, several authors \cite{roberts2002langevin, xifara2013langevin} have discussed these algorithms without considering them to be `geometric', rather simply Metropolis--Hastings methods in which proposal kernels have a position-dependent covariance structure.  We do not claim that the geometric view is the only one that should be taken.  Our goal is merely to point out that such position-dependent methods can often be viewed as methods defined on a manifold, and that studying the structure of the manifold itself may lead to new insights on the methods.  For example, taking the geometric viewpoint and noting the connection with Information geometry enabled Girolami \& Calderhead to adopt the Fisher metric for calculations \cite{girolami2011riemann}.  We list here a few open questions that the geometric viewpoint may help shed some insight on.

Computationally minded readers will have noted that using position-dependent covariance matrices adds a significant computational overhead in practice, with additional $O(n^3)$ matrix inversions required at each step of the corresponding Metropolis--Hastings algorithms.  Clearly there will be many problems for which the matrix $G(x)$ does not change very much, and therefore choosing a constant covariance $G^{-1}(x) = \Sigma$ may result in a more efficient algorithm overall.  Geometrically, this would correspond to a manifold with low curvature.  It may be that geometric ideas could be used to understand whether the manifold is \emph{flat enough} that a constant choice of $G(x)$ is sufficient.  To make sense of this truly would require a relationship between curvature, an inherently local property, and more global statements about the manifold.  Many results in differential geometry, beginning with the celebrated Gauss-Bonnet theorem, have previously related global and local properties in this way \cite{do1976differential}.  It is unknown to the authors whether results exist relating the curvature of a manifold to some global property, but this is an interesting avenue for further research.

A related question is when to choose the simplified manifold MALA over the full method.  Problems in which the term $\|\Lambda(x)\|$ is sufficient large to warrant calculation correspond to those for which the manifold has very high curvature in many places, so again making some global statement related to curvature could help here.

Although there is a reasonable intuitive argument for why the Hessian is an appropriate starting point for $G(x)$, the lack of positive-definiteness may be seen as a cause for concern by some.  After all, it could be argued that if the curvature is not positive-definite in a region, then how can it be a reasonable approximation to the local covariance structure.  Indeed, for target densities of the form $\pi(x) \propto e^{-|x|}$, the Hessian is everywhere equal to zero!  Much work in Information geometry has centred on the geometry of Hessian structures \cite{shima2007geometry}, and some insights from this field may help better understand the question of what appropriate metric to choose is.

Some recent work in high-dimensional inference has centred on defining MCMC methods for which efficiency scales $O(1)$ with respect to the dimension $n$ of $\pi(\cdot)$ \cite{beskos2013advanced, cotter2013mcmc}.  In the case where $X$ takes values in some infinite-dimensional function space, this can be done provided a Gaussian prior measure is defined for $X$.  A striking result from infinite-dimensional probability spaces is that two different probability measures defined over some infinite dimensional space have a striking tendency to have disjoint supports \cite{da2008stochastic}.  The key challenge for MCMC is to define transition kernels for which proposed moves are inside the support for $\pi(\cdot)$.  A straight-forward approach is to define proposals for which the prior is invariant, since the likelihood contribution to the posterior typically will not alter its support from that of the prior \cite{beskos2013advanced}.  However, the posterior may still look very different from the prior, as noted in \cite{law2013proposals}, so this proposal mechanism, though $O(1)$, can still result in slow exploration.  Understanding the geometry of the support, and defining methods which incorporate the likelihood term but also respect this geometry so as to ensure proposals remain in the support of $\pi(\cdot)$, is an intriguing research proposition.

The methods reviewed in this paper are based on first order Langevin diffusions.  Algorithms have also been developed which are based on \emph{second order} Langevin diffusions, in which a stochastic differential equation governs the behaviour of the velocity of a process \cite{ottobre2013function, horowitz1991generalized}.  A natural extension to the work of Girolami \& Calderhead \cite{girolami2011riemann} and Xifara et al. \cite{xifara2013langevin} would be to map such diffusions onto a manifold and derive Metropolis--Hastings proposal kernels based on the resulting dynamics.  The resulting scheme would be a generalisation of \cite{horowitz1991generalized}, though the most appropriate discretisation scheme for a second order process to facilitate sampling is unclear, and perhaps a question worthy of further exploration.

We have focused primarily here on the sample space $\mathcal{X} = \mathbb{R}^n$, and on defining an appropriate manifold on which to construct Markov chains.  In some inference problems, however, the sample space is a pre-defined manifold, for example the set of $n \times n$ rotation matrices, commonly found in the field of Directional Statistics \cite{mardia2009directional}.  Such manifolds are often not globally mappable to Euclidean $n$-space.  Methods have been devised for sampling from such spaces \cite{byrne2013geodesic, diaconis2013sampling}.  In order to use the methods described here for such problems, an appropriate approach for switching between coordinate patches at the relevant time would need to be devised, which could be an interesting area of further study.

Alongside these geometric problems, we can also discuss geometric MCMC methods from a statistical perspective.  The last example given in the previous section hinted that the manifold MALA may cope better with target distributions with heavy tails.  In fact, Latuszynski et al. \cite{latuszynski2011riemann} have shown that in one dimension, the manifold MALA is geometrically ergodic for a class of targets of the form $\pi(x) \propto \exp( -|x|^\beta )$ for any choice of $\beta \neq 1$.  This incorporates cases where tails are heavier than exponential and lighter than Gaussian, two scenarios under which geometric ergodicity fails for the MALA.

Finding optimal acceptance rates and scaling of $\lambda$ with dimension are two other related challenges.  In this case the picture is more complex.  Traditional results have been shown for Metropolis--Hastings methods in the case where target distributions are independent and identically-distributed, or some other suitable symmetry and regularity in the shape of $\pi(\cdot)$.  Manifold methods are, however, specifically tailored to scenarios in which this is \emph{not} the case, scenarios in which there is high correlation between components of $x$ which changes depending on the value of $x$.  It is less clear how to proceed with finding \emph{relevant} results which can serve as guidelines to practitioners here.  Indeed, Sherlock \cite{sherlock2013optimal} notes that a requirement for optimal acceptance rate results for the RWM to be appropriate is that the curvature of $\pi(x)$ does not change too much, yet this is the very scenario in which we would want to use a manifold method.




\section{Conclusions}

We have discussed the merits of viewing the sample space of a statistical model as a Riemannian manifold for Markov chain Monte Carlo.  We focused on the Metropolis-adjusted Langevin algorithm, and provided a full exposition of the relevant Markov chain and Riemannian geometry background in order to understand the method.  After this we provided a full geometric derivation of a Langevin diffusion on a Riemannian manifold, and resulting Metropolis--Hastings algorithms.  In the previous section, we have highlighted several open questions related to the emerging field of geometric Markov chain Monte Carlo, which we hope will inspire innovative new research in the field.


\section{Acknowledgements}

S. Livingstone is funded by a PhD Scholarship from Xerox Research Centre Europe.  M. Girolami is funded by an EPSRC Established Career Research Fellowship, EP/J016934/1 and a Royal Society Wolfson Research Merit Award.  The authors thank Michael Epstein and Simon Byrne for proofreading the article and giving useful suggestions.


\appendix

\section{Vector fields and the Covariant Derivative} \label{app:A}

Here we provide a short introduction to vector fields and differentiation on a smooth manifold, see \cite{boothby1986introduction, lee2003smooth}.  The following geometric notation is used here: (i) vector \emph{components} are indexed with a superscript, e.g. $v = (v^1,...,v^n)$, and (ii) repeated subscript and superscripts are summed over, e.g. $v^i\mathbf{e}_i = \sum_i v^i\mathbf{e}_i$ (known as the Einstein summation convention).

For any smooth manifold $M$, the set of all tangent vectors to points on $M$ is known as the \emph{tangent bundle}, and denoted $TM$.

A $C^r$ \emph{vector field} defined on $M$ is a mapping which assigns to each point $p \in M$ a tangent vector $v(p) \in T_pM$.  In addition, the components of $v(p)$ in any basis for $T_pM$ must also be $C^r$ \cite{boothby1986introduction}.  We will denote the set of all vector fields on $M$ as $\Gamma(TM)$.  For some vector field $v \in \Gamma(TM)$, at any point $p \in M$ the vector $v(p) \in T_pM$ can be written as a linear combination of some $n$ basis vectors $\{ \mathbf{e}_1,...,\mathbf{e}_n\}$ as $v = v^i\mathbf{e}_i$.  To understand how $v$ will change in a particular direction along $M$, it only makes sense therefore to consider derivatives along vectors in $T_pM$.  Two other things must be considered when defining a derivative along a manifold: (i) how the components $v^i$ of each basis vector will change, and (ii) how each basis vector $\mathbf{e}_i$ itself will change.  For the usual directional derivative on $\mathbb{R}^n$ the basis vectors do not change, as the tangent space is the same at each point, but for a more general manifold this is no longer the case: the $\mathbf{e}_i$'s are referred to as a `local' basis for each $T_pM$.

The covariant derivative $D^c$ is defined so as to account for these shortcomings.  When considering differentiation along a vector $u^* \notin T_pM$, $u^*$ is simply projected onto the tangent space.  The derivative with respect to any $u \in T_pM$ can now be decomposed into a linear combination of derivatives of basis vectors and vector components
\begin{equation} \label{eqn:covder1}
D^c_u[v] = D^c_{u^i\mathbf{e}_i}[v^i\mathbf{e}_i],
\end{equation} 
where the argument $p$ has been dropped but is implied for both components and local basis vectors.  The operator $D^c_u[v]$ is defined to be linear in both $u$ and $v$ and satisfy the product rule \cite{boothby1986introduction}, so (\ref{eqn:covder1}) can be decomposed into
\begin{equation} \label{eqn:covder2}
D^c_u[v] = u^i \left( D^c_{\mathbf{e}_i} [v^j]\mathbf{e}_j + v^j D^c_{\mathbf{e}_i}[\mathbf{e}_j] \right).
\end{equation}
The operator $D^c$ need therefore only be defined along the direction of basis vectors $\mathbf{e}_i$ and for vector component $v^i$ and basis vector $\mathbf{e}_i$ arguments.

For components $v^i$, $D^c_{\mathbf{e_j}}[v^i]$ is defined as simply the partial derivative $\partial_j v^i := \partial v^i/\partial x^j$.  The directional derivative of some basis vector $\mathbf{e}_i$ along some $\mathbf{e}_j$ is best understood through the example of a regular surface $\Sigma \subset \mathbb{R}^3$.  Here $D_{\mathbf{e}_j}[\mathbf{e}_i]$ will be a vector $w \in \mathbb{R}^3$.  Taking the basis for this space at the point $p$ as $\{\mathbf{e}_1,\mathbf{e}_2,\mathbf{\hat{n}}\}$, where $\mathbf{\hat{n}}$ denotes the unit normal to $T_p\Sigma$, we can write $w = \alpha\mathbf{e}_1 + \beta\mathbf{e}_2 + \kappa\mathbf{\hat{n}}$.  The covariant derivative $D^c_{\mathbf{e}_j}[\mathbf{e}_i]$ is simply the projection of $w$ onto $T_p\Sigma$, given by $w^* = \alpha\mathbf{e}_1 + \beta\mathbf{e}_2$.  More generally at some point $p$ in a smooth manifold $M$ the covariant derivative $D^c_{\mathbf{e}_j}[\mathbf{e}_i] = \Gamma^k_{ji}\mathbf{e}_k$ (with upper and lower indices summed over).  The coefficients $\Gamma^k_{ji}$ are known as the \emph{Christoffel symbols}: $\Gamma^k_{ji}$ denotes the coefficient of the $k$th basis vector when taking the derivative of the $i$th with respect to the $j$th.  If a Riemannian metric $g$ is chosen for $M$, then they can be expressed completely as a function of $g$ (or in local coordinates as a function of the matrix $G$).  Using these definitions, (\ref{eqn:covder2}) can be re-written as
\begin{equation}
D^c_u[v] = u^i \left( \partial_i v^k + v^j\Gamma^k_{ij} \right)\mathbf{e}_k.
\end{equation}

The divergence of a vector field $v \in \Gamma(TM)$ at the point $p \in M$ is given by
\begin{equation}
\text{div}_M(v) = D^c_{\mathbf{e}_i}[v^i],
\end{equation}
where again repeated indices are summed over.  If $M = \mathbb{R}^n$, this reduces to the usual sum of partial derivatives $\partial_i v^i$.  On a more general manifold $M$, the equivalent expression is
\begin{equation} \label{eqn:appdiv1}
D^c_{\mathbf{e}_i}[v^i] = \partial_i v^i + v^i\Gamma^j_{ij},
\end{equation}
where again repeated indices are summed.  As has been previously stated, if a metric $g$ and coordinate chart is chosen for $M$, the Christoffel symbols can be written in terms of the matrix $G(x)$.  In this case \cite{bernard1984geometrical}
\begin{equation}
\Gamma^j_{ij} = |G(x)|^{-\frac{1}{2}} \partial_i \left( |G(x)|^{\frac{1}{2}} \right),
\end{equation}
so (\ref{eqn:appdiv1}) becomes
\begin{equation}
D^c_{\mathbf{e}_i}[v^i] = |G(x)|^{-\frac{1}{2}} \partial_i \left( |G(x)|^{\frac{1}{2}} v^i \right),
\end{equation}
where $v = v(x)$.






\bibliography{entropy}
\bibliographystyle{unsrt}

\end{document}